\documentclass[prl,twocolumn,showpacs,preprintnumbers,amsmath,amssymb]{revtex4}
\usepackage[dvips]{graphicx}
\usepackage{graphicx}% Include figure files
\usepackage{amsmath}
\usepackage{amsbsy}
\usepackage{dcolumn}% Align table columns on decimal point
\usepackage{bm}% bold math

\begin{document}

\title{Superconducting gap variations induced by structural supermodulation in BSCCO}

\author{Brian M. Andersen$^{1,2}$ and P. J. Hirschfeld$^1$}

\affiliation{$^1$Department of Physics, University of Florida,
Gainesville, Florida 32611, USA\\
$^2$Nano-Science Center, Niels Bohr Institute, University of
Copenhagen, Universitetsparken 5, DK-2100 Copenhagen, Denmark}

\author{James A. Slezak}

\affiliation{LASSP, Department of Physics, Cornell University,
Ithaca, New York 14850, USA}

\date{\today}

\begin{abstract}

We discuss the possibility that the strain field introduced by the
structural supermodulation in Bi-2212 and certain other cuprate
materials may modulate the superconducting pairing interaction. We
calculate the amplitude of this effect, visible in scanning
tunneling spectroscopy experiments, and thereby relate a change in
the local superconducting gap with the change in the local dopant
displacements induced by the supermodulation. In principle, since
this modulation is periodic, sufficiently accurate x-ray
measurements or ab initio calculations should enable one to
determine which atomic displacements enhance pairing and therefore
$T_c$.

\end{abstract}


\maketitle

In the twenty years since the discovery of the high-$T_c$ cuprates,
an intense theoretical and experimental effort has not reached a
consensus on the origin of the phenomenon, although much has been
learned. Progress has been slow due both to the complexity of the
materials, and the difficulty of the theoretical problem of strong
electronic correlations. One advantage researchers in the high-$T_c$
field have is a set of new experimental tools which provide {\it
local} information on the electronic state. Chief among these new
methods is scanning tunneling spectroscopy (STS), which is providing
fascinating new insights and forcing new ways of thinking about the
high-$T_c$ problem by yielding an unprecedented level of
detail\cite{OEFischer:2006,cren,davisinhom1,Kapitulnik1,davisinhom2}.
Traditionally, such atomic scale information has been considered
irrelevant to the phenomenon of superconductivity, since rapid
oscillations of pair wavefunctions on length scales smaller than the
coherence length $\xi$ are integrated out in the conventional
pairing theory, where it is assumed that $\xi\gg a$, with  $a$ the
lattice spacing.  In the cuprate superconductors, on the other hand,
$\xi$ is much smaller than in conventional materials, approaching
$a$ itself, and it is conceivable that a new type of approach
accounting for atomic scale physics will be required to solve this
problem.

This possibility was highlighted recently when McElroy {\sl et al.}
\cite{KMcElroy:2005} discovered the positive correlation of the
positions of dopant atoms in superconducting Bi-2212 with the local
energy gap. It was argued that these atoms were in fact interstitial
O atoms\cite{KMcElroy:2005}, probably located between the BiO and
SrO layers\cite{YHe:2006}. Subsequently, Nunner {\sl et
al.}\cite{TSNunner:2005a} showed that a good fit to several
independent measured STS correlations could be understood if one
assumed that the dopants, in addition to delivering holes to the
CuO$_2$ plane and inducing a screened Coulomb potential, also
modulate the local pair interactions. The idea is that each dopant
distorts the lattice around it in such a way as to modify the
effective electronic structure, characterized by hopping integrals
$t$, $t'$, ..., exchange constants $J$, electron-phonon couplings
$\lambda$, resulting in a modified local pairing interaction between
electrons as well\cite{JXZhu:2005,Maska:2007}. Phenomenological fits
to the data of McElroy {\sl et al.}\cite{KMcElroy:2005} then led to
the conclusion that a substantial modulation of the pair interaction
on an atomic scale is present in the effective low-energy
Hamiltonian of the disordered BSCCO
material\cite{TSNunner:2005a,TSNunner:2005b,BMAndersen:2006,TSNunner:2006}.

More recently, Slezak {\sl et al.}\cite{JASlezak:2007} performed an
extensive STS study of the supermodulation (SM) in BSCCO. From these
local measurements it was found that the gap is modulated with the
SM phase $\phi^{SM}$ with an amplitude of order 10\% of its average
value in near-optimally doped samples, providing a remarkable direct
quantitative link between atomic displacements in the unit cell and
the local superconducting gap $\Delta$. In other words, the SM
induces a pair-density wave in Bi-2212.

In this work we extend the model of Ref.
\onlinecite{TSNunner:2005a} to include effects of the SM strain
field. In a similar spirit, we assume that the local atomic
displacements caused by the SM generate an additional, sinusoidal
modulation of $g$, which we refer to as a $g$-wave.  The
supermodulation presumably originates from a mismatch between the
insulating layers and the preferred bond lengths of perovskite
crystal. It can be characterized by a wavevector $q_{SM}$, has a
wavelength of approximately $\lambda_{SM}=2\pi/q_{SM} \approx 26
\AA \approx 4.8~\mbox{unit cells}$, runs along the $a$-axis, and
leads to displacements of the atomic positions of up to 0.4 \AA.
It is believed that one of the main effects of the SM is to
modulate the distance between the apical oxygen and the CuO$_2$
plane\cite{kanmoss,gao,grebille}. For simplicity, we neglect the
periodic changes in other terms in the Hamiltonian, such as the
electron hopping. We will show that the amplitude of this $g$-wave
can in principle be determined by comparison with experiment,
given the assumptions already specified. In this case, one should
be able to relate the change in the pair potential $\delta g$ to
the atomic displacements caused by the SM, information which
should be available with sufficiently precise x-ray data, or from
ab initio calculations.   The philosophy here is similar to that
of Nunner {\sl et al.}\cite{TSNunner:2005a}, but the measurement
of changes in the local gap caused by the supermodulation has the
advantage that the associated periodic displacements should be
easier to determine empirically than in the case of the random O
dopants. As discussed below, current x-ray data on Bi-2212 are not
yet able to answer this question, but there appears to be no
fundamental obstacle to improving the precision so as to be able
to achieve this goal; they may then be able to remove any
remaining ambiguity as to the microscopic origin of the modulated
pairing in this material.

In the following,  we use the conventional $d$-wave BCS Hamiltonian
defined on a 2D square lattice
\begin{equation}
\label{eq:hamiltonian} \hat{H}\!=\!\sum_{{\langle ij
\rangle}\sigma}\! t_{ij} \hat{c}_{i\sigma}^\dagger
\hat{c}_{j\sigma} +\! \sum_{i\sigma} \!
(V_i-\mu)\hat{c}_{i\sigma}^\dagger \hat{c}_{i\sigma} \!+\!
\sum_{\langle ij \rangle} \! \left( \Delta_{ij}
\hat{c}_{i\uparrow}^\dagger \hat{c}_{j\downarrow}^\dagger \!+\!
\mbox{H.c.} \! \right)\!,\!
\end{equation}
where in the first term we include nearest $t$ and next-nearest
$t^\prime=-0.3t$ neighbor hopping.   We set the chemical potential
$\mu=-t$ in order to model the Fermi surface of BSCCO near optimal
doping, and $\sum_{\langle ij \rangle}$ denotes summation over
neighboring lattice sites $i$ and $j$. Disorder of the usual
screened Coulomb type is included in the impurity potential $V_i
=V_0 f_i$ where $f_i=\sum_s \exp(-r_{is}/\lambda)/r_{is}$, where
$r_{is}$ is the distance from a defect $s$ to the lattice site $i$
in the plane. Distance (energy) is measured in units of $\sqrt{2}a$
($t$), where $a$ is the Cu-Cu distance, and the calculations are
done at $T=0$. Note that the particular form of $f_i$ is merely a
convenient way to vary the range of the impurity potential landscape
through the parameter $\lambda$. The $d$-wave order parameter
$\Delta_{ij}=g_{ij} \langle \hat{c}_{i\uparrow}
\hat{c}_{j\downarrow} - \hat{c}_{j\downarrow}
\hat{c}_{i\uparrow}\rangle/2$ is determined self-consistently via
\begin{equation}
\Delta_{ij}=\frac{g_{ij}}{2} \sum_n \left( u_n(i) v_n(j) + v_n(i)
u_n(j)\right) \tanh(\frac{E_n}{2 T}),
\end{equation}
Here, $\{ E_n , u_n , v_n\}$ is the eigensystem resulting from
diagonalization of the Bogoliubov-de Gennes (BdG) equations
associated with Eq.(\ref{eq:hamiltonian}). The pairing interaction
$g_{ij}$ is assumed to vary in space relative to its homogeneous
value $g_0$ as
\begin{equation}
g_{ij}=g_0+\delta g_{imp} (f_i+f_j)/2 + \delta g_{SM} \cos
(\phi^{SM}_{i}),\label{eqn:g}
\end{equation}
where $\delta g_{imp}$ and $\delta g_{SM}$ are the amplitudes of
the dopant and $g$-wave modulation, respectively, and $i,j$ are
nearest neighbors. In the homogeneous case $g_{ij}=g_0$ we choose
$g_0=1.16t$ giving $\Delta_{ij}=0.1t$. In the inhomogeneous case
we make sure to adjust $g_0$, in order to maintain the same
average gap as in the corresponding homogeneous system. The third
($g$-wave) term in Eq. (\ref{eqn:g}) is taken to vary sinusoidally
with the SM phase $\phi^{SM}_i$, a phase variable associated with
the structural supermodulation as determined in Ref.
\onlinecite{JASlezak:2007}. In the following, we use the
experimentally determined $\phi^{SM}_i$ as input in Eq.
(\ref{eqn:g}), and ignore for simplicity the conventional
potential ($\tau_3$ channel) associated with each dopant, and
consider disorder only in the pair ($\tau_1$) channel. The goal
will be to adjust the amplitude $\delta g_{SM}$ to obtain better
quantitative agreement with the experiments of Ref.
\onlinecite{JASlezak:2007}.

\begin{figure}[b]
\begin{minipage}{.49\columnwidth}
\hspace{-0.5cm}(a)\\
\vspace{-0.1cm}\includegraphics[clip=true,bb=242 148 650
485,width=.95\columnwidth]{mapdiluted_lowres.epsf}
\end{minipage}
\begin{minipage}{.49\columnwidth}
\hspace{-0.5cm}(b)\\
\vspace{-0.1cm}\includegraphics[clip=true,bb=242 148 650
485,width=.95\columnwidth]{41026A00GapMap_gp.epsf}
\end{minipage}\\
\begin{minipage}{.49\columnwidth}
\hspace{-0.5cm}(c)\\
\includegraphics[clip=true,bb=242 148 650
485,width=.95\columnwidth]{newImpPot90Vi0001lam05rz05.epsf}
\end{minipage}
\begin{minipage}{.49\columnwidth}
\hspace{-0.5cm}(d)\\
\includegraphics[clip=true,bb=242 148 650
485,width=.95\columnwidth]{GapMap_lowres.epsf}
\end{minipage}
\begin{minipage}{.47\columnwidth}
\vspace{+0.1cm}
\hspace{-0.5cm}(e)\\
\hspace{-1.2cm}\includegraphics[clip=true,width=.90\columnwidth]{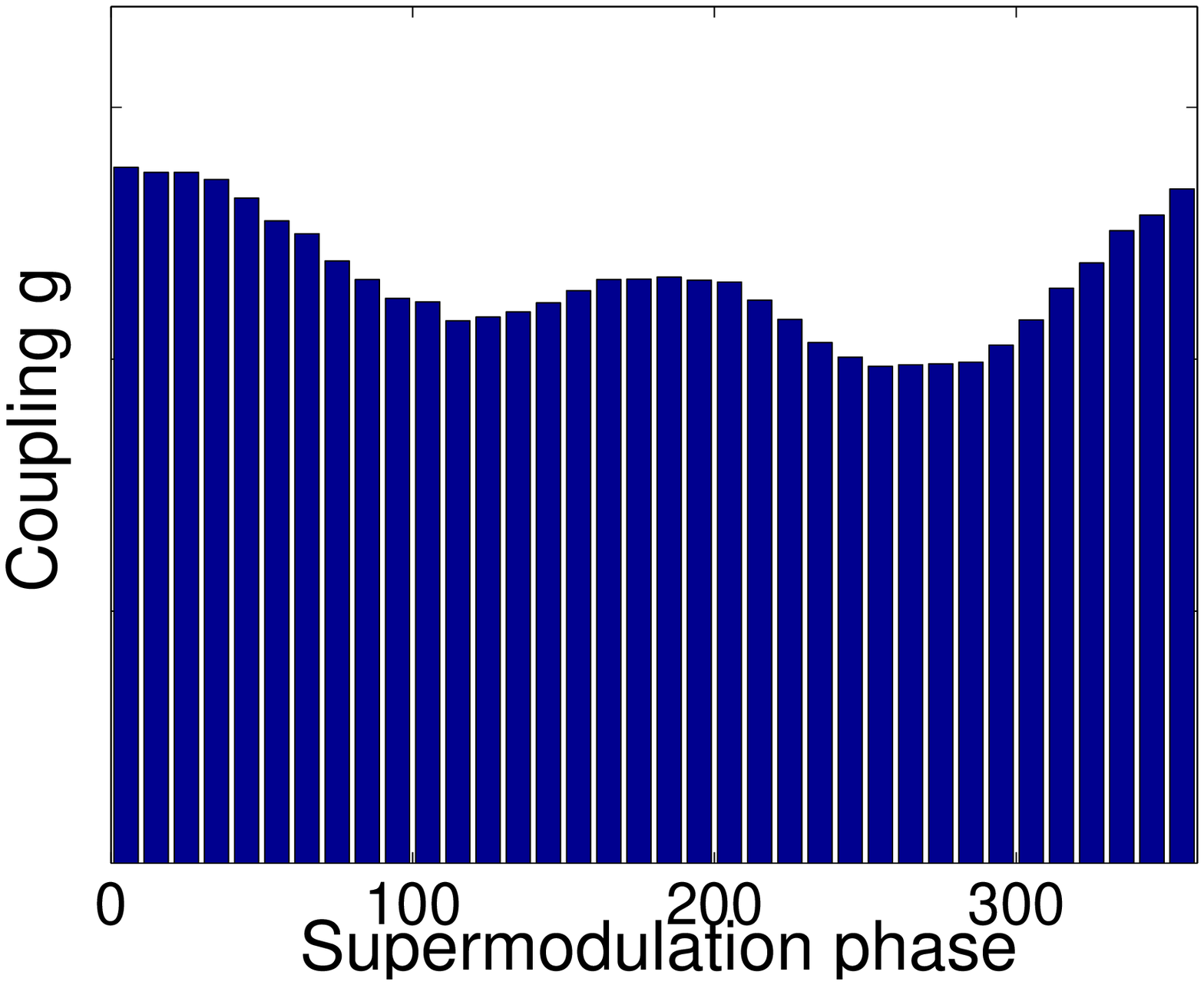}
\end{minipage}
\begin{minipage}{.47\columnwidth}
\vspace{+0.1cm}
\hspace{-0.3cm}(f)\\
\hspace{-1.0cm}\includegraphics[clip=true,width=.95\columnwidth]{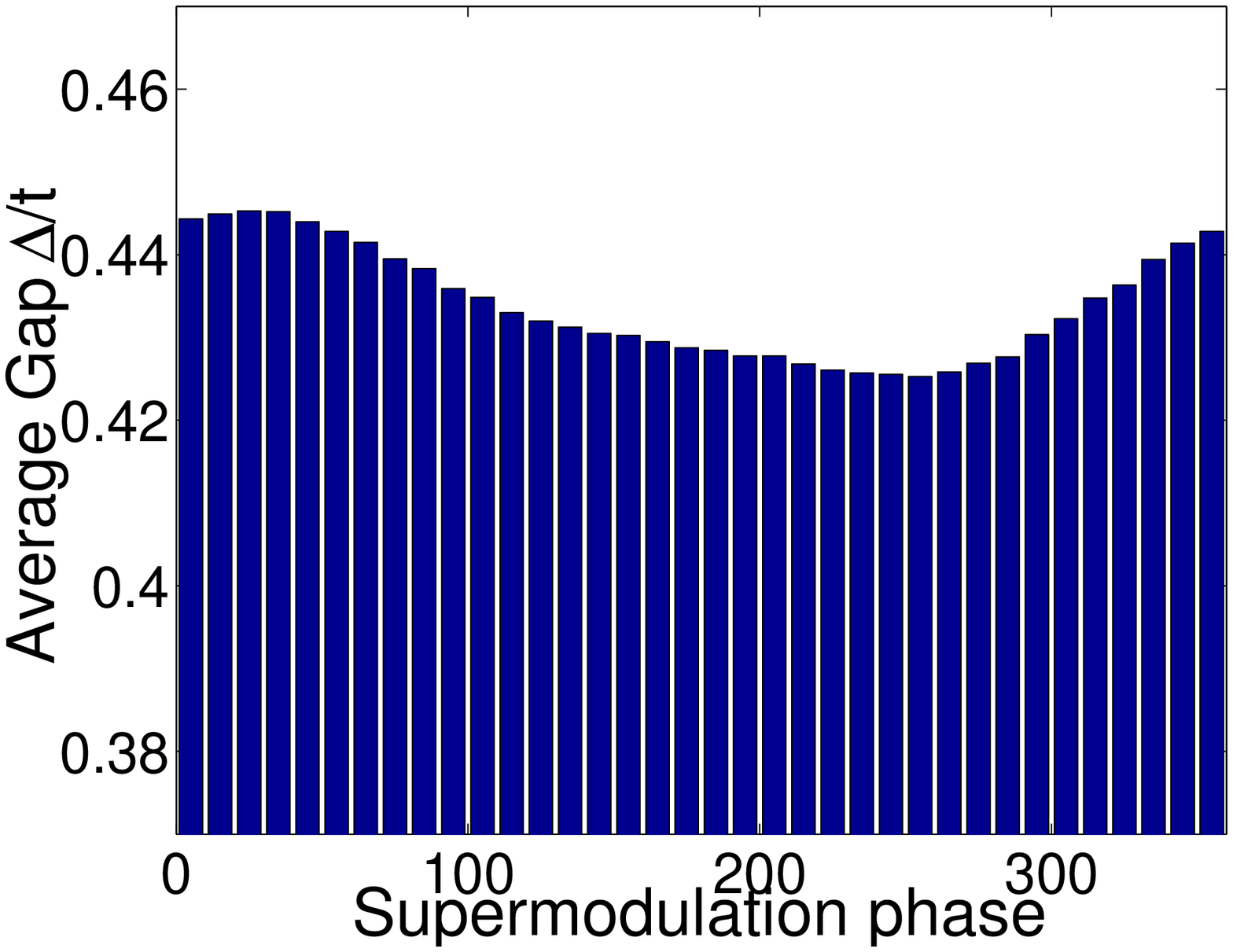}
\end{minipage}
\caption{(Color online) (a) Experimental $dI/dV$ map [arb. units]
at -960 meV of an optimally doped BSCCO
sample\cite{KMcElroy:2005}. (b) Experimental gapmap [meV] in the
same region as (a) at $T=4$K. (c) The theoretical impurity
potential extracted from (a) assuming a distance from the CuO$_2$
plane $r_z=0.5$ and $\lambda=0.5$. (d) The gapmap resulting from
using (c) as the pairing potential in the BdG equations with
$\delta g_{imp}=1.5t$. (e) and (f) display the $g$-histogram and
$\Delta$-histogram versus $\phi^{SM}$, respectively. The model
results (c-f) are for $\delta g_{SM}=0.0$.} \label{fig:2Dmaps}
\end{figure}

That a nonzero value of $\delta g_{SM}$ is required is illustrated
in Fig. \ref{fig:2Dmaps} which shows results obtained with $\delta
g_{SM}=0$\cite{BMAndersen:2006}. The input to the theory is, in a
field of view (FOV) of $49\mbox{nm}\times 49\mbox{nm}$, the
measured conductance at -960 meV (a), known to track the positions
of the dopant oxygens.  The corresponding gapmap obtained by
McElroy {\sl et al.}\cite{KMcElroy:2005} is shown in (b). In Fig.
\ref{fig:2Dmaps}(c) we display the impurity potential generated by
assuming that each of these dopants provides a potential centered
on the bright spot positions of (a) which decays as a screened
exponential {\it in the pairing channel}, as described above. The
similarity to (a) is manifest. The experimental FOV is modeled as
a $90\times 90$ lattice system rotated 45 degrees with respect to
the 3.83 \AA $~$  Cu-Cu bonds. Therefore our system consists of
$2\times 90\times 90$ sites and is aligned with the experimental
FOV. The gapmap computed from the coherence peak-to-peak distance
in the theoretical local density of states (LDOS) using (c) as
input pair potential $\delta g_{imp}$ is shown Fig.
\ref{fig:2Dmaps}(d). Gapmaps reasonably consistent with experiment
are found for $\delta g_{imp} \sim 1.5t$.

\begin{figure}[b]
\begin{minipage}{.49\columnwidth}
\hspace{+0.3cm}\vspace{-0.2cm}(a)\\
\includegraphics[clip=true,bb=120 200 500 660,width=.85\columnwidth,angle=270]{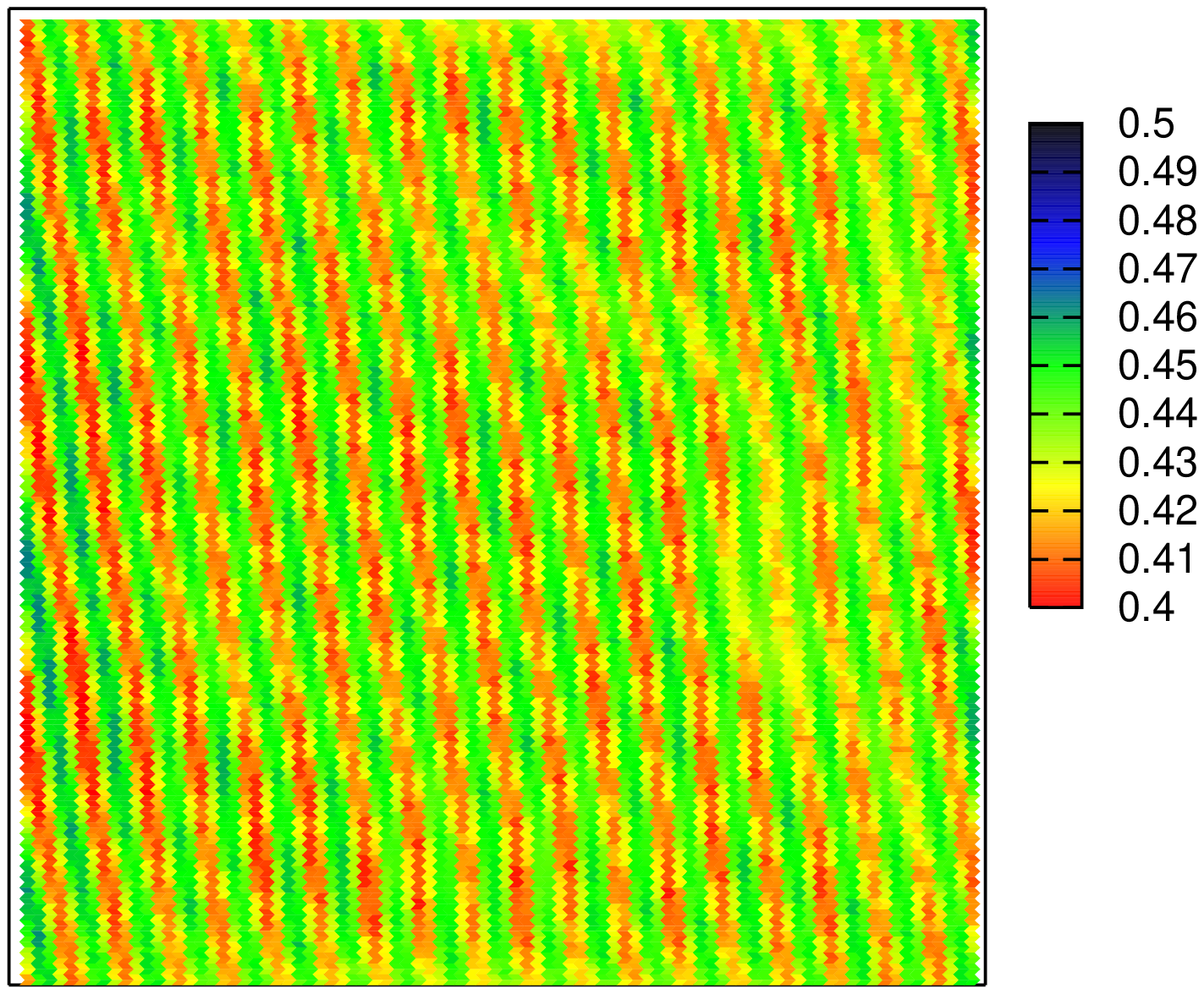}
\end{minipage}
\begin{minipage}{.47\columnwidth}
\hspace{+0.3cm}(b)\\
\includegraphics[clip=true,width=.95\columnwidth]{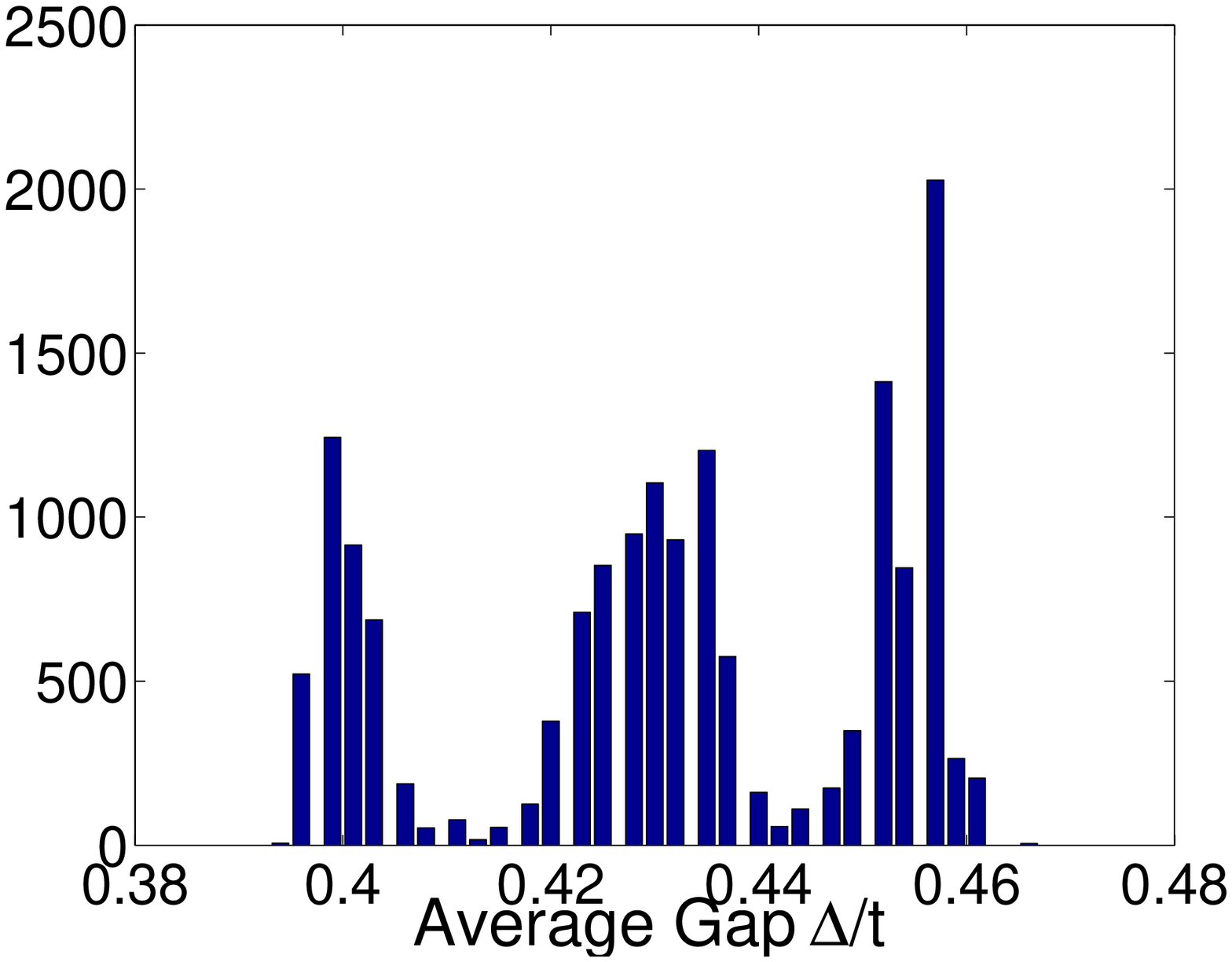}
\end{minipage}\\
\begin{minipage}{.47\columnwidth}
\hspace{+0.3cm}(c)\\
\hspace{-0.5cm}\includegraphics[clip=true,width=.95\columnwidth]{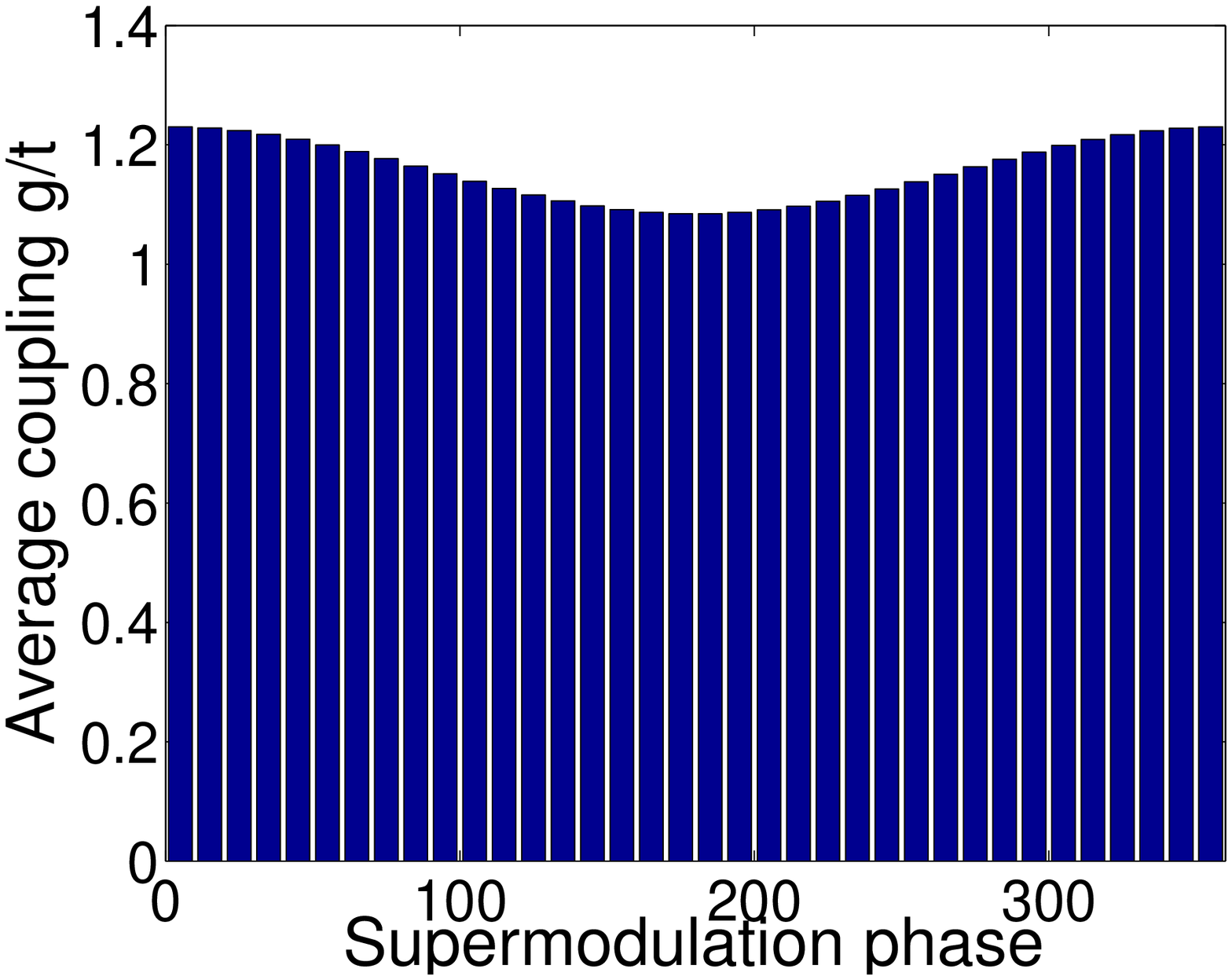}
\end{minipage}
\begin{minipage}{.47\columnwidth}
\hspace{+0.5cm}(d)\\
\hspace{-0.1cm}\includegraphics[clip=true,width=.95\columnwidth]{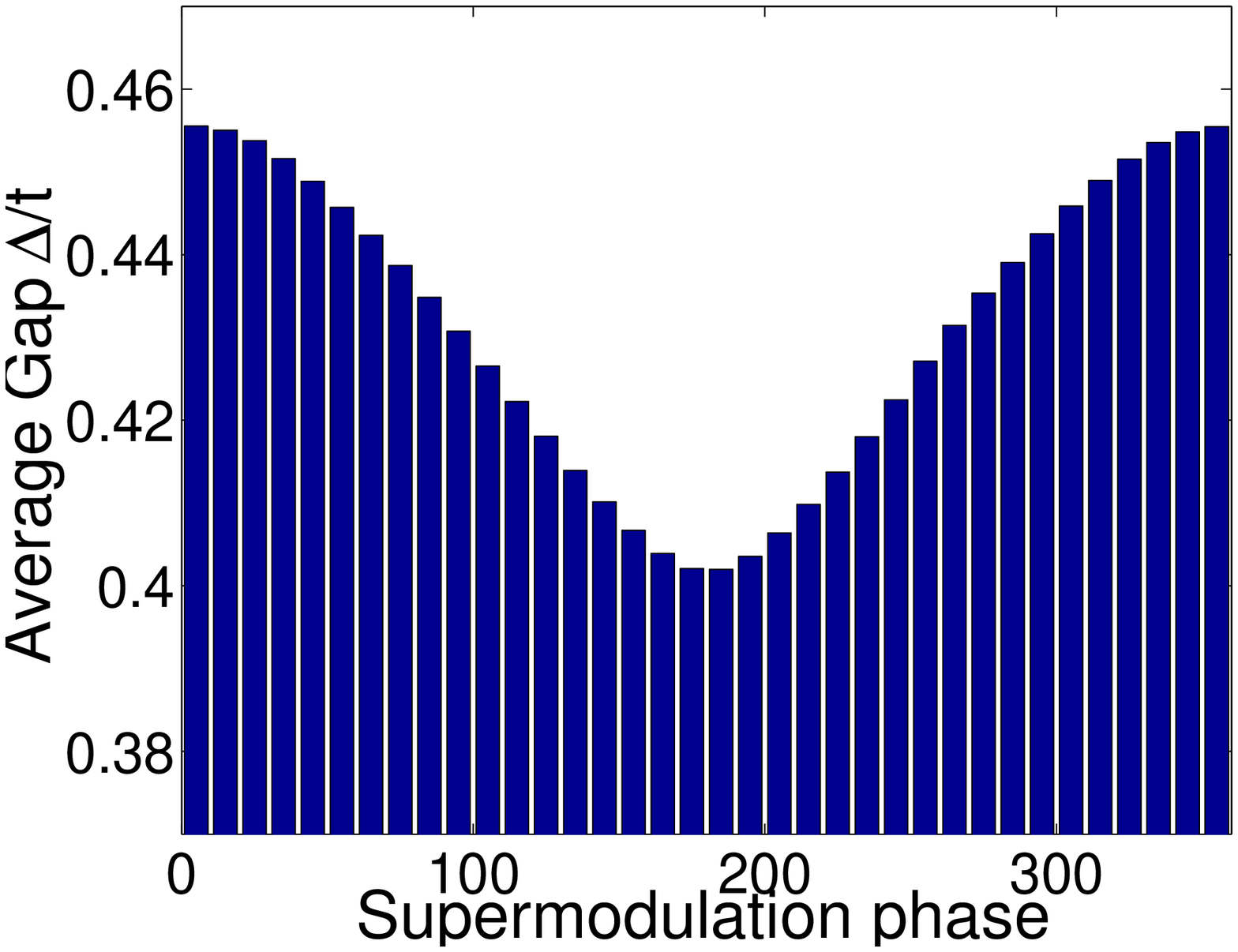}
\end{minipage}
\caption{(Color online) For clarity we show here the
self-consistent results for the superconducting gap $\Delta$
resulting from a pure $g$-wave ($\delta g_{imp}=0$ and $\delta
g_{SM}=0.08t)$: (a) real-space gapmap, (b) conventional gap
histogram, (c) $g$-histogram versus $\phi^{SM}$, and (d)
$\Delta$-histogram versus $\phi^{SM}$.} \label{fig:pureg}
\end{figure}

\begin{figure}[t]
\begin{minipage}{.49\columnwidth}
\vspace{-0.4cm}(a)\\
\includegraphics[clip=true,bb=120 200 500 660,width=.85\columnwidth,angle=270]{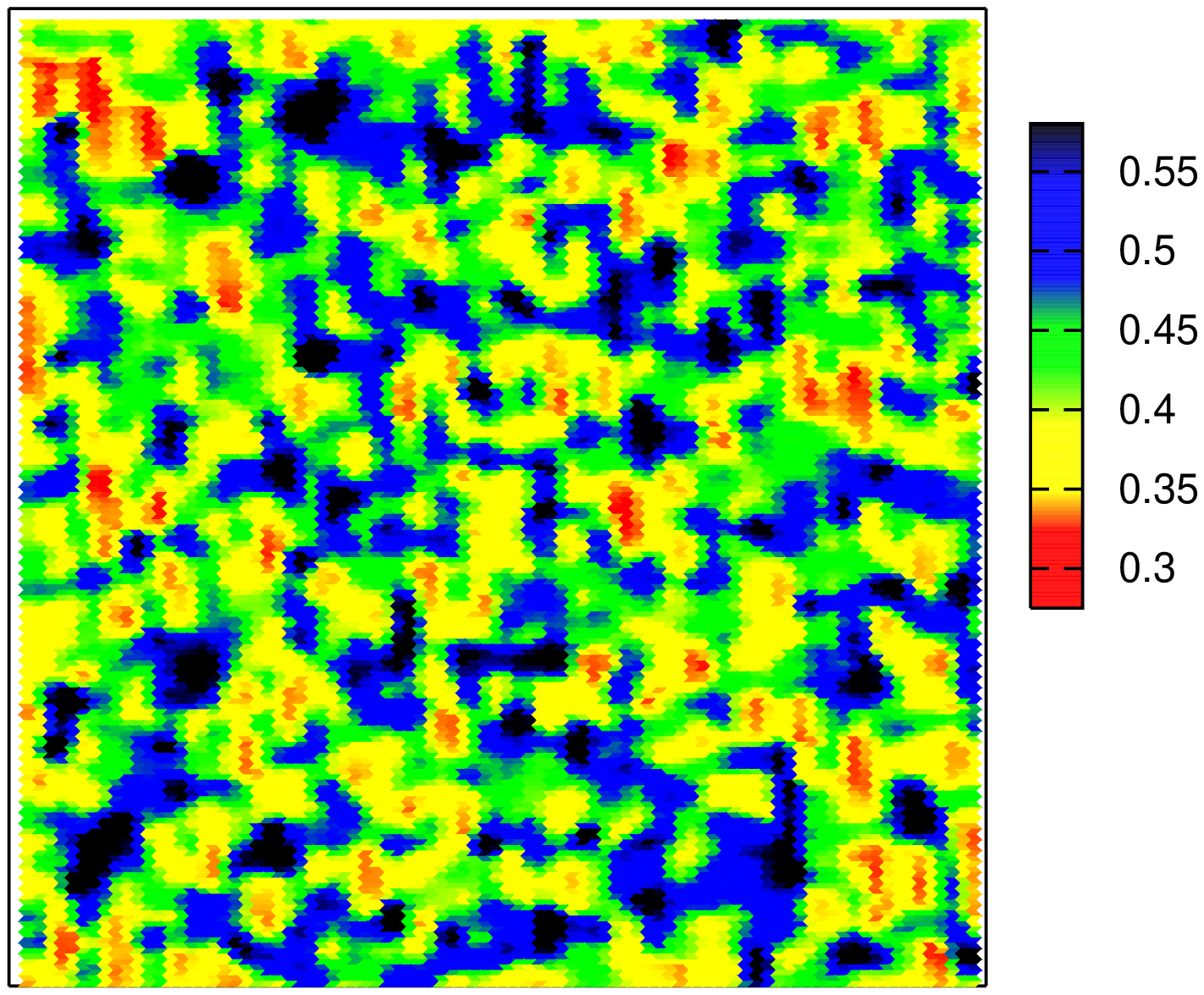}
\end{minipage}
\begin{minipage}{.49\columnwidth}
\vspace{-0.4cm}(b)\\
\includegraphics[clip=true,bb=120 200 500 660,width=.85\columnwidth,angle=270]{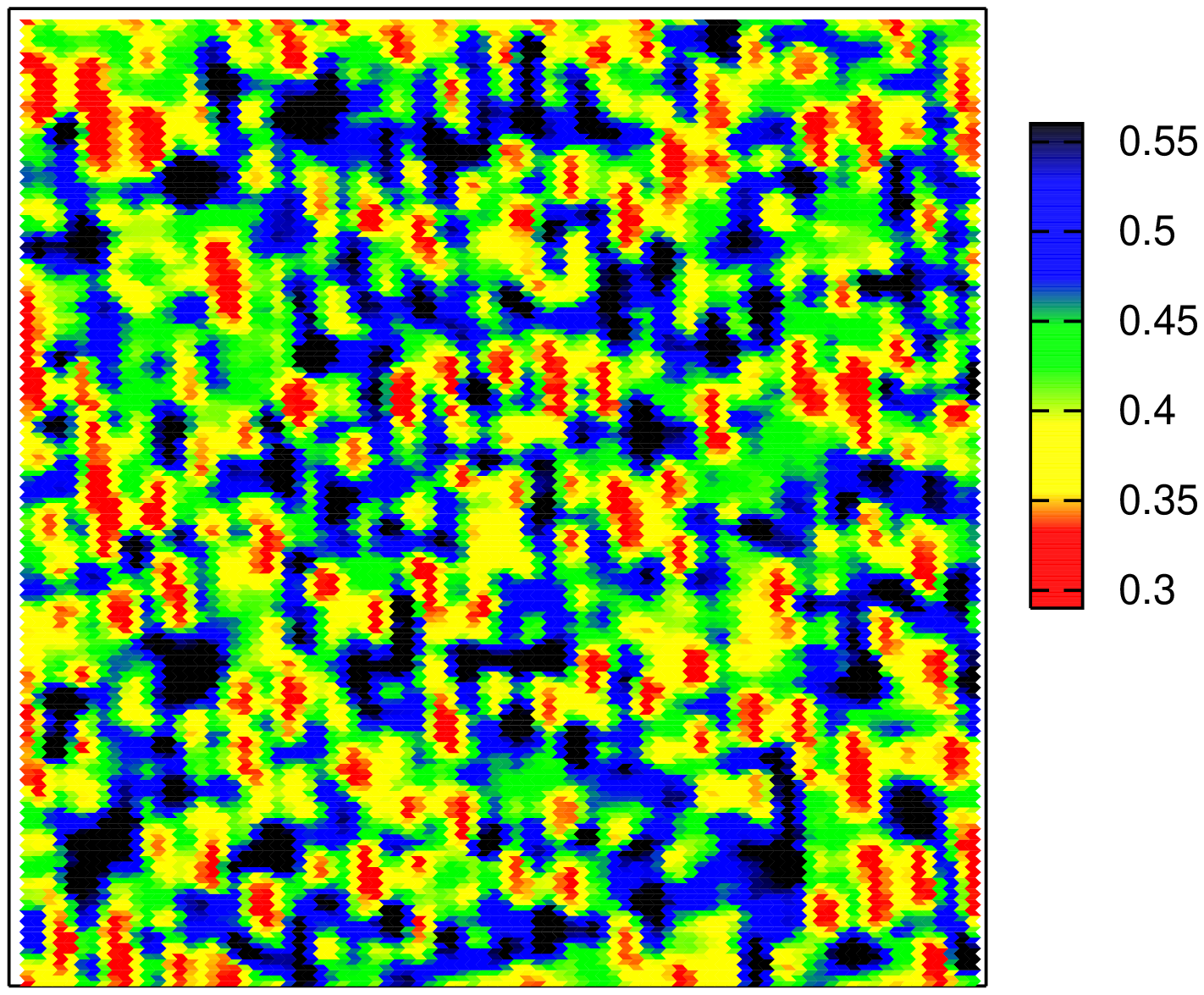}
\end{minipage}\\
\begin{minipage}{.47\columnwidth}
(c)\\
\hspace{-0.8cm}\includegraphics[clip=true,width=.95\columnwidth]{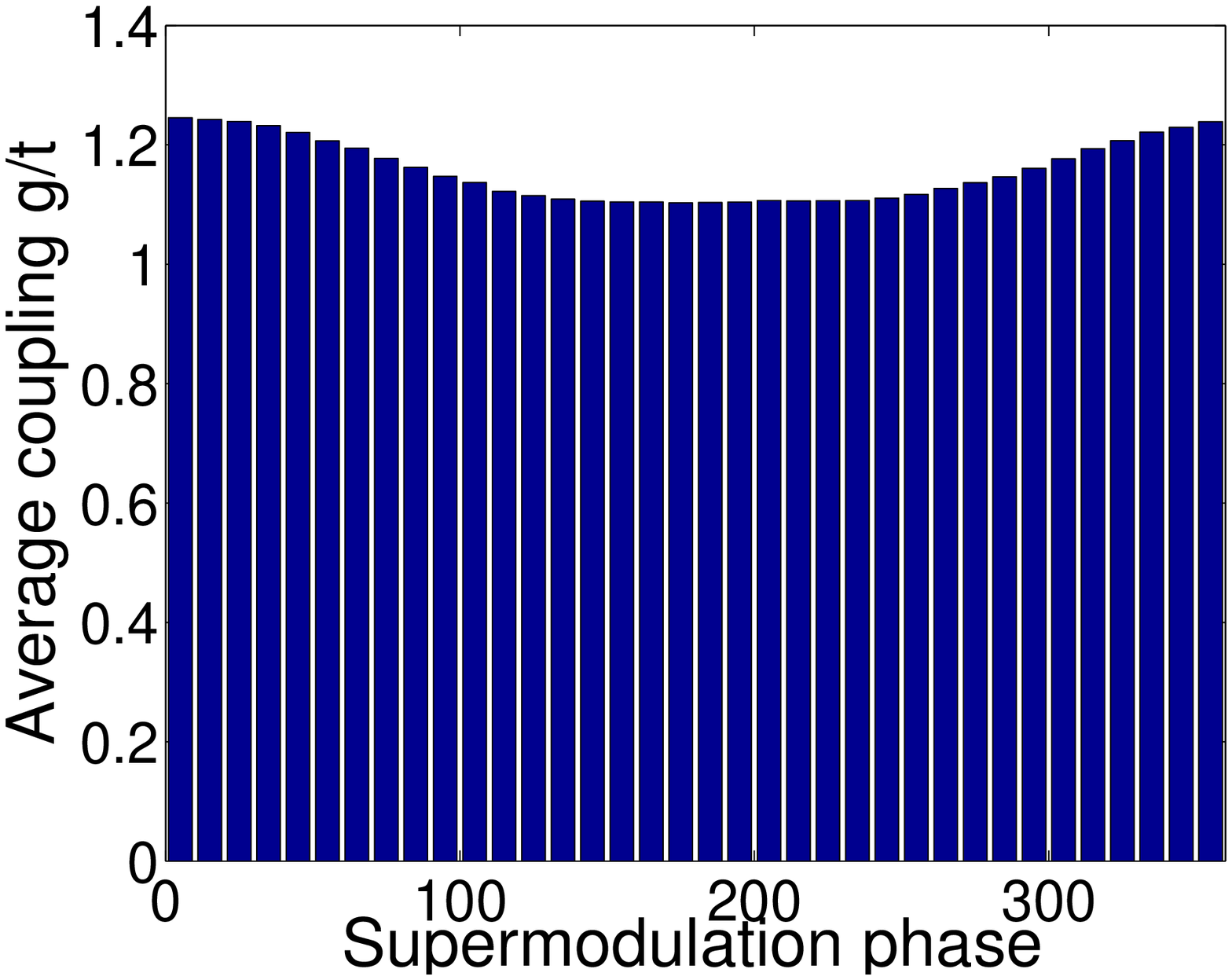}
\end{minipage}
\begin{minipage}{.47\columnwidth}
(d)\\
\hspace{-0.6cm}\includegraphics[clip=true,width=.95\columnwidth]{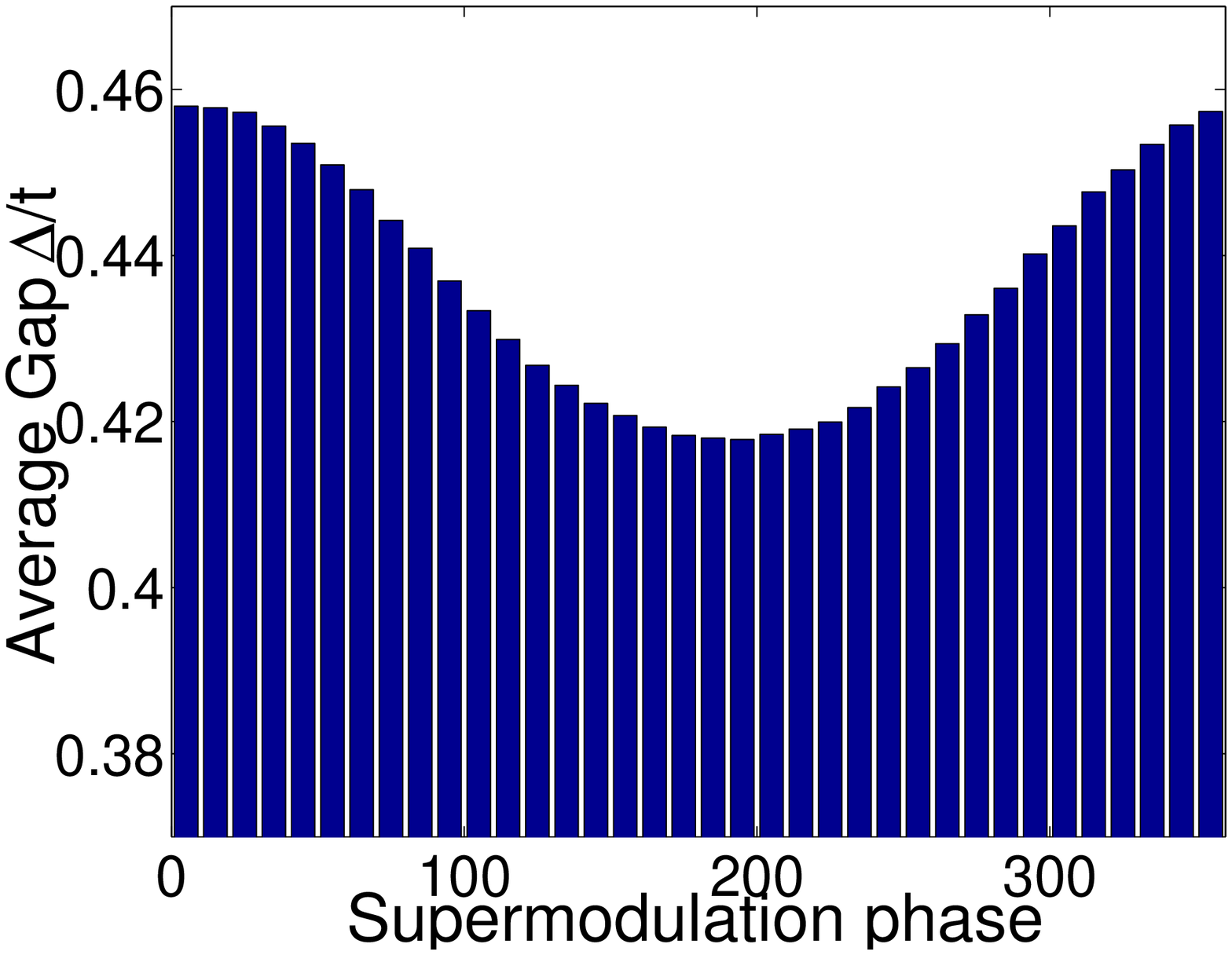}
\end{minipage}
\caption{(Color online) (a,b) Real-space gapmaps for $\delta
g_{imp}=1.5t$, and $\delta g_{SM}=0.06t$ (a) and $\delta
g_{SM}=0.14t$ (b). For the parameters in (a) we show is (c) the
$g$-histogram versus $\phi^{SM}$, and in (d) the $\Delta$-histogram
versus $\phi^{SM}$.\label{gwaveanddopants}}
\end{figure}

Although various correlations among O positions and the LDOS
$\rho(E)$ are successfully reproduced by the Nunner {\sl et al.}
\cite{TSNunner:2005a} approach, one deficiency appears upon closer
examination of the gapmaps Fig. \ref{fig:2Dmaps}(b) and (d). The
experimental result (Fig. \ref{fig:2Dmaps}(b)) contain nearly
vertical linear striated modulations visible to the eye, which match
the corresponding STM topograph and are therefore caused by the SM.
These are not obviously manifest in the theoretical gapmap presented
in Fig. \ref{fig:2Dmaps} (d). In fact, weak correlations of the
dopant positions with the supermodulation phase are indeed present
in the data, as seen in Fig. \ref{fig:2Dmaps}(e-f) and lead, for the
disorder parameters which reproduced the gapmap statistics, to a
small net modulation of the gap with $\phi^{SM}$.

In Fig. \ref{fig:2Dmaps}(e), it is seen that the experimental dopant
distribution is correlated with $\phi^{SM}$ at 0 and 180$^\circ$,
leading to a  peaks in the coupling $g$ vs $\phi^{SM}$ at these
phases in our model. Fig. \ref{fig:2Dmaps}(f) shows a histogram of
the gap in Fig. \ref{fig:2Dmaps}(d) versus SM phase $\phi^{SM}$.
Because of the effective smearing of the order parameter response
over scales of order the coherence length $\xi\sim\lambda_{SM}$, the
180$^\circ$ peak is partially wiped out and a resulting weak
360$^\circ$-periodic modulation of 4-5\% of $\Delta$ versus
$\phi^{SM}$ remains, qualitatively similar to the
experiments\cite{JASlezak:2007}. Increasing the phenomenological
amplitude of the dopant potential leads to overall fluctuations of
the gap amplitude which are much too large compared to experiment.
Therefore,  in order to generate a 10\% variation of $\Delta$ versus
$\phi^{SM}$, and maintain quantitative agreement with the spatial
extent and overall amplitude of the gap variations evident from the
experimental gapmaps, we need to include a nonzero $\delta g_{SM}$.
This is fully consistent with the notion\cite{TSNunner:2005a} that
the reason the oxygens modulate the pairing is due to local
distortions of the lattice, and the SM should produce similar
effects on its own.  Hence, in the following we investigate the
possibility that a $g$-wave is present in the system in addition to
the dopant $\tau_1$ disorder.

In Fig. \ref{fig:pureg}, we first show the effects of a pure
$g$-wave without $\tau_1$ disorder, i.e. $\delta g_{imp}=0$. From
Fig. \ref{fig:pureg}(a), which displays the gap in real-space, it is
seen that the SM agrees well to that observed in the corresponding
experimental FOV by Slezak {\sl et al.}\cite{JASlezak:2007} as it
should per construction. The gap histogram is shown in Fig.
\ref{fig:pureg}(b), and in Fig. \ref{fig:pureg}(c) (Fig.
\ref{fig:pureg}(d)) we show the histograms for the input $g$-wave
(self-consistent $\Delta$) versus SM phase $\phi^{SM}$, both of
which trivially exhibit a sinusoidal dependence. Fig.
\ref{gwaveanddopants} displays typical results when including both
$\delta g_{imp}$ and $\delta g_{SM}$ into the simulation. Here, we
used the same disorder parameters as in Andersen {\sl et
al.}\cite{BMAndersen:2006} to obtain the gapmaps (Fig.
\ref{fig:2Dmaps}(c,d)), but with $\delta g_{SM}=0.06t$ (a) and
$\delta g_{SM}=0.14t$ (b). In both these plots the striated gap
modulations are evident. For the parameters used in Fig.
\ref{gwaveanddopants}(a), we show in \ref{gwaveanddopants}(c) and
\ref{gwaveanddopants}(d) the g-histogram and $\Delta$-histogram,
respectively. As seen, the oscillation of $g$ versus $\phi^{SM}$ of
roughly sinusoidal form is still found, with amplitude close to the
input $\delta g_{SM}$. The gap response has a corresponding
amplitude of about 10\% of its average value  in agreement with the
measurements by Slezak {\sl et al.}\cite{JASlezak:2007}. Therefore,
within the modulated pairing scenario, an experimental gap
modulation amplitude of approximately 10\%, is caused by a
SM-induced $g$-wave of similar size relative to the background
pairing strength $g_0$. Note that after including the SM phase, it
is Fig. \ref{gwaveanddopants}(a) that should be compared to the
experimental gapmap in Fig. \ref{fig:2Dmaps}(b).

We would now like to extract which type of atomic displacements
are present in the SM and associated with the enhanced pairing at
0 and 360 degrees SM phase. In principle, this information should
be available from careful x-ray diffraction data, but this is
complicated by the existence of deviations of this system from
stoichiometry, both through Bi/Sr substitutions and oxygen
interstitials, which together determine the incommensurability of
the true system.  Early x-ray
analysis\cite{VPetricek:1990,AFMarshall:1988,YLePage:1989,JMTarascon:1988}
suggested a weak correlation of the interstitial oxygen position
with the SM, but as remarked above this correlation does not
appear to be the most significant one present in the STM gapmaps.

More recently, progress in the analysis of incommensurate
systems\cite{AALevin:1994,HFFan:1999} has been made, and new
short-range ordered structures have been
identified\cite{MIzquierdo:2006}, but the various studies disagree
at essential points\cite{JASlezak:2007}, and it does not appear to
be possible with present data to identify the actual displacements
of atoms near the SM maximum or minimum with sufficient precision
to eventually draw conclusions regarding the microscopic origin of
the pairing modulations. There does appear to be a statistically
significant correlation of the $z$ coordinate of the apical oxygen
relative to the CuO$_2$ plane with the SM
phase\cite{AYamamoto:1990}, but it anti-correlates with the local
gap\cite{YOhta:1991}. This behavior is the opposite of that which
might be expected on the basis of the analysis by Pavarini {\it et
al.}\cite{EPavarini:2001}, who pointed out an empirical
correlation between apical O coordinate $z$ and $T_c$. This may
point to the primacy of other atomic displacements, or suggest
that $T_c$ itself is less related to in-plane pairing strength and
more to interlayer couplings. In any case, our analysis should
provide incentive for a repeated assault on the x-ray analysis of
this compound, with the hope of eventually providing direct local
information on the origin of the pairing.

{\it Acknowledgements.} The authors are grateful to J. C. Davis for
advice and encouragement at all stages of this work. Calculations
were performed at the   University of Florida High-Performance
Computing Center (http://hpc.ufl.edu).


\begin{thebibliography}{99}
%
\bibitem{OEFischer:2006} \O. Fischer, M. Kugler, I. Maggio-Aprile, C. Berthod,
and C. Renner, Rev. Mod. Phys. {\bf 79}, 353 (2007).
%
\bibitem{cren} T. Cren, D. Roditchev, W. Sacks, J. Klein, J.-B. Moussy, C. Deville-Cavellin, and M. Lagu\"{e}s, Phys. Rev.
Lett. {\bf 84}, 147 (2000).
%
\bibitem{davisinhom1} S.-H. Pan, P. O'Neal, R. L. Badzey, C. Chamon, H. Ding, J. R. Engelbrecht, Z. Wang,
H. Eisaki, S. Uchida, A. K. Gupta, K.-W. Ng, E. W. Hudson, K. M.
Lang, and J. C. Davis, Nature {\bf 413}, 282 (2001).
%
\bibitem {Kapitulnik1} C. Howald, P. Fournier, and A. Kapitulnik,
Phys. Rev. B {\bf 64}, 100504(R) (2001).
%
\bibitem{davisinhom2} K. M. Lang, V. Madhavan, J. E. Hoffman, E. W. Hudson, H. Eisaki, S. Uchida, and J. C. Davis
Nature {\bf 415}, 412 (2002).
%
\bibitem{KMcElroy:2005} K. McElroy, H. Eisaki, S. Uchida, and J. C. Davis,
Science {\bf 309}, 1048 (2005).
%
\bibitem{YHe:2006} Y. He, T. S. Nunner, P. J. Hirschfeld, and H.-P. Cheng, Phys. Rev. Lett. {\bf 96}, 197002 (2006).
%
\bibitem{TSNunner:2005a} T. S. Nunner, B. M. Andersen, A. Melikyan, and P. J.
Hirschfeld, Phys. Rev. Lett. {\bf 95}, 177003 (2005).
%
\bibitem{JXZhu:2005} J.-X. Zhu, cond-mat/0508646.
%
\bibitem{Maska:2007} M. M. Ma\'{s}ka, \.{Z}. \'{S}led\'{z}, K. Czajka, and M.
Mierzejewski, cond-mat/0703566.
%
\bibitem{TSNunner:2005b} T. S. Nunner, W. Chen, B. M. Andersen, A. Melikyan, and P. J. Hirschfeld, Phys. Rev. B {\bf 73}, 104511 (2006).
%
\bibitem{BMAndersen:2006}B. M. Andersen, A. Melikyan, T. S. Nunner, and P. J. Hirschfeld, Phys. Rev. B {\bf 74}, 060501(R) (2006).
%
\bibitem{TSNunner:2006} T. S. Nunner, P. J. Hirschfeld, B. M. Andersen, A. Melikyan, K.
McElroy, cond-mat/0606685.
%
\bibitem{JASlezak:2007} J. A. Slezak, Ph.D. thesis, Cornell University 2007.  J. A. Slezak {\sl et al.}, (unpublished).
%
\bibitem{kanmoss} X. B. Kan and S. C. Moss,
Acta. Cryst. B {\bf 48}, 122 (1991).
%
\bibitem{gao} Y. Gao, P. Lee, P. Coppens, M. A. Subramanian, and A. W. Sleight,
Science {\bf 241}, 954 (1988).
%
\bibitem{grebille} D. Grebille, H. Leligny, A. Ruyter, P. Labbe, and B. Raveau, Acta Cryst. B {\bf 52}, 628 (1996).
%
\bibitem{VPetricek:1990} V. Petricek, Y. Gao, P. Lee, and P.
Coppens, Phys. Rev. B {\bf 42}, 387 (1990).
%
\bibitem{AFMarshall:1988} A. F. Marshall, B. Oh, S. Spielman, M. Lee, C. B. Eom, R. W. Barton, R. H. Hammond, A. Kapitulnik, M.R. Beasley, and T. H.
Geballe, Appl. Phys. Lett. {\bf 53}, 426 (1988).
%
\bibitem{YLePage:1989} Y. Le Page, W. R. McKinnon, J. M. Tarascon, and P.
Barboux, Phys. Rev. B {\bf 40}, 6810 (1989).
%
\bibitem{JMTarascon:1988} J. M. Tarascon, Y. Le Page, P. Barboux, B. G.
Bagley, L. H. Greene, W. R. McKinnon, G. W. Hull, M. Giroud, and D.
M. Hwang, Phys. Rev. B {\bf 37}, 9382 (1988).

\bibitem{AALevin:1994} A. A. Levin, Yu. I Smolin and Yu. F.
Shepelev, J. Phys. Cond. Mat {\bf 6}, 3539 (1994).
%
\bibitem{HFFan:1999} H. F. Fan, Microscopy Research and Technique
{\bf 46}, 104 (1999).
%
\bibitem{MIzquierdo:2006} M. Izquierdo, S. Megtert, J. P. Albouy, J. Avila, M. A. Valbuena, G. Gu, J. S. Abell, G. Yang, M. C.
Asensio, and R. Comes, Phys. Rev. B {\bf 74}, 054512 (2006).

\bibitem{AYamamoto:1990} A. Yamamoto, M. Onoda, E.  Takayama-Muromachi, F. Izumi, T. Ishigaki, and H. Asano, Phys. Rev. B {\bf 42},
4228 (1990).
%
\bibitem{YOhta:1991} Y. Ohta, T. Tohyama, and S. Maekawa, Phys. Rev. B {\bf 43},
2968 (1991).
%
\bibitem{EPavarini:2001} E. Pavarini, I. Dasgupta, T. Saha-Dasgupta, O. Jepsen, and O. K.
Andersen, Phys. Rev. Lett. {\bf 87}, 047003 (2001).
\end{thebibliography}
\end{document}